\documentclass[11pt]{article}
\usepackage[body={16cm,23cm}]{geometry}
\usepackage{nccfoots}

\usepackage[hang,small,center]{caption}
\usepackage{amsmath,amssymb}
\usepackage{amsfonts}
\usepackage{epic}
\usepackage{eepic}
\usepackage{graphicx}
\usepackage{cite}
\newcommand{\ds}{\displaystyle}

\renewcommand{\author}[1]{\large\rm #1\\ \bigskip}
\newcommand{\address}[1]{{\normalsize\it #1\\}\bigskip}
\renewcommand{\title}[1]{\bigskip\bigskip\Large\bf #1\bigskip\bigskip\\}
\newcommand{\keywords}[1]{\begin{quote}\small\sl Keywords: #1\end{quote}}

\newcommand{\Bigpsi}[3]{\phantom{\Psi}_2 \kern -.05em
\Psi_2\left(\genfrac{}{}{0pt}{}{#1}{#2}\biggl|#3\right)}

\newcommand{\be}{\begin{equation}}
\newcommand{\ee}{\end{equation}}

\newcommand{\x}{{\boldsymbol{x}}}

\newcommand{\T}{{\mathsf T}}
\newcommand{\alg}{{\cal{A}}}
\newcommand{\R}{{\cal{R}}}
\newcommand{\ii}{\mathsf{i}}
\newcommand{\kop}{\boldsymbol{k}}
\newcommand{\bos}{\boldsymbol{a}}

\newcommand{\opr}{\stackrel{\textrm{def}}{=}}
\newcommand{\cdil}{\varphi}
\def\EXP{\textrm{{\large e}}}

\begin{document}

\vglue 2cm
\begin{center}

\title{QUANTUM GEOMETRY OF 3-DIMENSIONAL LATTICES\\
AND TETRAHEDRON EQUATION\Footnote{$\dagger$}{%
Plenary talk at the 
XVI International Congress on Mathematical Physics, 
3-8 August 2009, Prague, Czech Republic}}
 
\author {Vladimir V. Bazhanov$^*$ and Vladimir V. Mangazeev}
\address{Department of Theoretical Physics,\\
         Research School of Physical Sciences and Engineering,\\
    Australian National University, Canberra, ACT 0200, Australia.\\
$^*$E-mail: Vladimir.Bazhanov@anu.edu.au}
\author{Sergey M. Sergeev}
\address{Faculty of Information Sciences and Engineering,\\
University of Canberra, Bruce ACT 2601, Australia.\\
E-mail: Sergey.Sergeev@canberra.edu.au}
\begin{abstract}
We study geometric consistency relations between angles of
3-dimensional (3D) circular quadrilateral lattices --- lattices whose
faces are planar quadrilaterals inscribable into a circle. We show
that these relations generate canonical transformations of a
remarkable ``ultra-local'' Poisson bracket algebra defined on
discrete 2D surfaces consisting of circular quadrilaterals.
Quantization of this structure allowed us to obtain new solutions of
the tetrahedron equation  (the 3D analog of the Yang-Baxter equation)
as well as reproduce all those that were previously known.
These solutions generate an infinite number of
non-trivial solutions of the Yang-Baxter equation and also define
integrable 3D models of statistical mechanics and quantum field theory.
The latter can be thought of as describing quantum fluctuations of
lattice geometry. 
\end{abstract}

\end{center}
\newpage
\vglue .3cm
\section{Introduction}
Quantum integrability is traditionally understood as
a {\em purely algebraic} phenomenon.
It stems from the Yang-Baxter equation \cite{Yang:1967,Bax72} and
other algebraic structures such as the
affine quantum groups \cite{Dri87,Jim85}
(also called the quantized Kac-Moody algebras), the Virasoro algebra
\cite{BPZ84} and their representation theory.
It is, therefore, quite interesting to learn that these algebraic
structures also have remarkable geometric origins \cite{BMS08a}, which will
be reviewed here. 

Our approach \cite{BMS08a} is based on connections between integrable {\em
three-dimensional} (3D) quantum systems and
integrable models of 3D discrete differential geometry.
The analog of the Yang-Baxter equation for integrable quantum  systems
in 3D is called the {\em tetrahedron equation}.
It was introduced by Zamolodchikov in
\cite{Zamolodchikov:1980rus,Zamolodchikov:1981kf}  (see also
\cite{Bazhanov:1981zm,Baxter:1986phd,Bazhanov:1992jq,
Bazhanov:1993j,Kashaev:1993ijmp,Korepanov:1993jsp,
KashaevKorepanovSergeev,BS05} for further
important results in this field).
Similarly to the Yang-Baxter equation the tetrahedron
equation provides local integrability conditions which are not
related to the size of the lattice. Therefore the same solution of
the tetrahedron equation defines different integrable models on
lattices of different size, e.g., for finite
periodic cubic lattices. Obviously, any such three-dimensional
model can be viewed as a two-dimensional integrable model on a
square lattice, where the additional third dimension is treated as
an internal degree of freedom. Therefore every solution of the
tetrahedron equation provides an infinite sequence of integrable
2D models differing by the size of this ``hidden third
dimension''\cite{Bazhanov:1981zm}. 
Then a natural question arises whether known 2D
integrable models can be obtained in this way.
Although a complete answer to this question is unknown, a few
non-trivial examples of such correspondence
have already been constructed. The first example \cite{Bazhanov:1992jq} 
reveals the 3D structure of 
the generalized  chiral Potts model \cite{BKMS,Date:1990bs}.
Another example \cite{BS05} reveals 3D structure of all
two-dimensional solvable models associated with 
finite-dimensional highest weight representations for 
quantized affine algebras $U_q(\widehat{sl}_n)$,
$n=2,3,\ldots,\infty$ (where $n$ coincides with the size
of the hidden dimension).

Here we unravel yet another remarkable property of the same
solutions of the tetrahedron equation (in addition to the hidden
3D structure of the Yang-Baxter equation and quantum groups). We show
that these solutions  
can be obtained from quantization of geometric
integrability conditions for the 3D {\em circular 
lattices} --- lattices whose faces are planar quadrilaterals 
inscribable into a circle.

The 3D circular lattices were introduced \cite{Bob96} as a
discretization of orthogonal coordinate systems, originating from classical
works of Lam\'e \cite{Lame1859} and Darboux \cite{Darboux}.
In the continuous case such coordinate systems are described by
integrable partial differential equations (they are connected with the
classical soliton theory \cite{ZakharovManakov,Kri96}). Likewise, the
quadrilateral and circular lattices are described by integrable 
difference equations. The key idea of the geometric approach
\cite{BobenkoPinkall,Bob96,DoliwaSantini,AdlerBobenkoSuris,
KonopelchenkoSchief,CDS97,KS98, DMS98, BoSurUMN07} 
to integrability of discrete classical systems is to
utilize various consistency conditions 
\cite{BoSurbook} arising from geometric relations
between elements of the lattice. It is quite remarkable that these conditions
ultimately reduce to certain 
incidence theorems of elementary geometry. 
For instance, the integrability conditions
for the quadrilateral lattices merely reflect the 
fact of existence of the 4D Euclidean cube \cite{DoliwaSantini}. 
In Sect.2 we present these conditions algebraically 
in a standard form of the {\em functional tetrahedron equation}
\cite{KashaevKorepanovSergeev}. The latter serves 
as the classical analog of the quantum tetrahedron equation, discussed
above, and
provides a connecting link to integrable quantum systems. 

In Sect.~\ref{poisson-sec} we study 
relations between edge angles on the 3D circular quadrilateral
lattices and show that these relations describe  symplectic transformations of
a remarkable ``ultra-local'' Poisson algebra on quadrilateral
surfaces (see Eq.\eqref{poisson}).
Quantization of this structure allows one
to obtain {\em   all currently known solutions} 
of the tetrahedron equation. They are presented
in Sect.~\ref{solutions}, namely\footnote{
The other known solutions, previously found by Hietarinta
\cite{Hietarinta:1994pt} and
Korepanov \cite{Korepanov:1993jsp}, 
were shown to be special cases of \cite{SMS96}.},  
\begin{itemize}
\item[(I)]
{Zamolodchikov-Bazhanov-Baxter solution \cite{Zamolodchikov:1981kf,
Bazhanov:1992jq,Kashaev:1993ijmp,SMS96}}   
\item[(II)]
{Bazhanov-Sergeev solution \cite{BS05}}
\item[(III)]
{Bazhanov-Mangazeev-Sergeev solution \cite{BMS08a}}
\end{itemize}
including their interaction-round-a-cube and vertex forms. 
Additional details on the corresponding solvable 3D models, in particular,
on their quasi-classical limit and connections with geometry, can be
found in \cite{BMS08a}.

\section{Discrete differential geometry: ``Existence as integrability''}
In this section we consider classical discrete integrable systems
associated with the quadrilateral lattices.
There are several ways to extract algebraic integrable systems
from the geometry of these lattices. One approach,
developed in
\cite{Kas96,BogdanovKonopelchenko, DoliwaSantini, DS00,
KonopelchenkoSchief},
leads to discrete analogs of the
Kadomtsev-Petviashvili integrable hierarchy.
Here we present a different approach exploiting the angle geometry
of the 3D quadrilateral lattices.

\subsection{Quadrilateral lattices}
Consider three-dimensional lattices, obtained by
embeddings of the integer cubic lattice ${\mathbb Z}^3$ into the
$N$-dimensional Euclidean space ${\mathbb R}^N$, with $N\ge 3$.
Let $\x({m})\in {\mathbb R}^N$, denote coordinates of the lattice
vertices, labeled by the 3-dimensional integer vector
 $m=m_1e_1+m_2e_2+m_3e_3\in
{\mathbb Z}^3$, where $e_1=(1,0,0), e_2=(0,1,0)$ and
$e_3=(0,0,1)$. Further, for any given lattice vertex $\x_0=\x(m)$,
the symbols $\x_i=\x(m+e_i)$, $\x_{ij}=\x(m+e_i+e_j)$, etc., will
denote neighboring lattice vertices. The lattice is called
{\em quadrilateral\/}\  if all its faces $(\x_0,\x_i,\x_j,\x_{ij})$
are planar quadrilaterals.
The existence of these lattices is based on the
following elementary geometry fact (see Fig.~1) \cite{DoliwaSantini},
\begin{figure}[ht]
\centering
\setlength{\unitlength}{0.3mm}
\begin{picture}(170,170)
\allinethickness{.6mm}
 \path(0,20)(100,0)(150,40)(140,130)(90,150)(20,120)(0,20)
 \path(20,120)(100,100)
 \path(100,0)(100,100)
 \path(140,130)(100,100)
\allinethickness{.3mm}
 \dashline{5}(0,20)(80,60)
 \dashline{5}(150,40)(80,60)
 \dashline{5}(90,150)(80,60)
 \put(-10,10){\scriptsize $\boldsymbol{x}_{23}$}
 \put(100,-10){\scriptsize $\boldsymbol{x}_{3}$}
 \put(150,30){\scriptsize $\boldsymbol{x}_{13}$}
 \put(143,135){\scriptsize $\boldsymbol{x}_{1}$}
 \put(87,155){\scriptsize $\boldsymbol{x}_{12}$}
 \put(7,120){\scriptsize $\boldsymbol{x}_{2}$}
 \put(77,50){\scriptsize $\boldsymbol{x}_{123}$}
 \put(105,95){\scriptsize $\boldsymbol{x}_{0}$}
%

\end{picture}
\caption{An elementary hexahedron of a cubic quadrilateral lattice.}
\label{fig-cube1}
\end{figure}
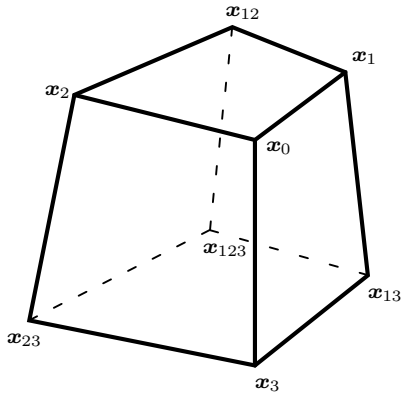
%
\begin{quote}
{\em Consider four points $\x_0, \x_1, \x_2, \x_3$
in general position in ${\mathbb R}^N$, $N\ge 3$.
On each of the three planes $(\x_0,\x_i,\x_j)$,
$1\le i <j\le 3$ choose an extra point $\x_{ij}$ not lying on the lines
$(\x_0,\x_i)$, $(\x_0,\x_j)$ and $(\x_i,\x_j)$. Then there exist a unique point
$\x_{123}$ which simultaneously belongs to the three planes
$(\x_1,\x_{12},\x_{13})$, $(\x_2,\x_{12},\x_{23})$ and
$(\x_3,\x_{13},\x_{23})$.}
\end{quote}

\noindent The six planes, referred to above, obviously lie in the
same 3D subspace of the target space. They define a hexahedron
with quadrilateral faces, shown in Fig.~\ref{fig-cube1}. It has
the topology of the cube, so we will call it ``cube'', for
brevity. Let us study elementary geometry relations among the
\begin{figure}[ht]
\centering
\setlength{\unitlength}{0.3mm}
\begin{picture}(400,170)
\put(230,10){\begin{picture}(150,150)
\allinethickness{.6mm}
 \path(0,20)(100,0)(150,40)(140,130)(90,150)(20,120)(0,20)
\thinlines
 \dashline{5}(0,20)(80,60)
 \dashline{5}(150,40)(80,60)
 \dashline{5}(90,150)(80,60)
 \put(-10,10){\scriptsize $\boldsymbol{x}_{23}$}
 \put(100,-10){\scriptsize $\boldsymbol{x}_{3}$}
 \put(150,30){\scriptsize $\boldsymbol{x}_{13}$}
 \put(143,135){\scriptsize $\boldsymbol{x}_{1}$}
 \put(87,155){\scriptsize $\boldsymbol{x}_{12}$}
 \put(7,120){\scriptsize $\boldsymbol{x}_{2}$}
 \put(17,21){\scriptsize $\beta_3'$}
 \put(7,32){\scriptsize $\delta_2'$}
 \put(69,64){\scriptsize $\gamma_2'$}
 \put(80,50){\scriptsize $\alpha_3'$}
 \put(85,65){\scriptsize $\delta_1'$}
 \put(22,112){\scriptsize $\beta_2'$}
 \put(78,138){\scriptsize $\alpha_2'$}
 \put(92,138){\scriptsize $\beta_1'$}
 \put(93,5){\scriptsize $\delta_3'$}
 \put(132,36){\scriptsize $\gamma_3'$}
 \put(137,47){\scriptsize $\gamma_1'$}
 \put(129,123){\scriptsize $\alpha_1'$}
 \put(-10,150){(b)}
\end{picture}}
\put(10,10){\begin{picture}(150,150)
\allinethickness{.6mm}
 \path(0,20)(100,0)(150,40)(140,130)(90,150)(20,120)(0,20)
 \path(20,120)(100,100)
 \path(100,0)(100,100)
 \path(140,130)(100,100)
\thinlines
 \put(-10,10){\scriptsize $\boldsymbol{x}_{23}$}
 \put(100,-10){\scriptsize $\boldsymbol{x}_{3}$}
 \put(150,30){\scriptsize $\boldsymbol{x}_{13}$}
 \put(143,135){\scriptsize $\boldsymbol{x}_{1}$}
 \put(87,155){\scriptsize $\boldsymbol{x}_{12}$}
 \put(7,120){\scriptsize $\boldsymbol{x}_{2}$}
 \put(5,23){\scriptsize $\delta_1$}
 \put(22,110){\scriptsize $\beta_1$}
 \put(90,7){\scriptsize $\gamma_1$}
 \put(88,93){\scriptsize $\alpha_1$}
 \put(103,13){\scriptsize $\delta_2$}
 \put(139,42){\scriptsize $\gamma_2$}
 \put(129,117){\scriptsize $\alpha_2$}
 \put(105,95){\scriptsize $\beta_2$}
 \put(95,107){\scriptsize $\delta_3$}
 \put(40,120){\scriptsize $\beta_3$}
 \put(123,127){\scriptsize $\gamma_3$}
 \put(86,141){\scriptsize $\alpha_3$}
 \put(-10,150){(a)}
\end{picture}}
\end{picture}
\caption{The ``front'' (a)  and ``back'' (b) faces of the cube in
Fig.~\ref{fig-cube1} and their angles.}\label{fig-cube2}
\end{figure}
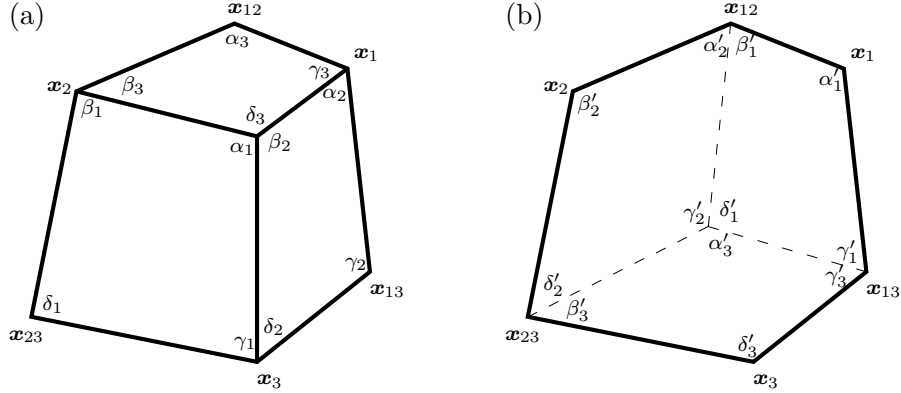
angles of this cube. Denote the angles between the edges as in
Fig.~\ref{fig-cube2}. Altogether we have $6 \times 4=24$ angles,
connected by six linear relations
\begin{equation}
\alpha_j+\beta_j+\gamma_j+\delta_j=2\pi ,\qquad
\alpha'_j+\beta'_j+\gamma'_j+\delta'_j=2\pi, \qquad
j=1,2,3,\label{2pi}
\end{equation}
which can be immediately solved for all ``$\delta$'s''. This
leaves 18 angles, but only nine of them are independent. 
Indeed, a mutual arrangement (up to an overall rotation)
of unit normal vectors to six planes 
in the 3D-space is determined by
nine angles only. 
Once this arrangement is fixed all other angles
can be calculated. Thus the nine independent angles of the three
``front'' faces of the cube, shown in Fig.\ref{fig-cube2}a,
completely determine the angles on the three ``back'' faces, shown
in Fig.\ref{fig-cube2}b, and vice versa. So the geometry of our
cube provides an invertible map for three triples of independent
variables
\begin{equation}
{\cal R}_{123}:\qquad \{\alpha_j,\beta_j,\gamma_j\}\to
\{\alpha'_j,\beta'_j,\gamma'_j\}, \qquad j=1,2,3.\label{map}
\end{equation}
Suppose now that all angles are known. To completely define the
cube one also needs  to specify lengths of its three edges. All
the remaining edges can be then determined from  simple linear
relations. Indeed, the four sides of every quadrilateral are
constrained by two relations, which can be conveniently presented
in the matrix form
\begin{equation}
\left(\begin{array}{c} \ell_p'\\[.1cm]
\ell_q'\end{array}\right)\;=\;
X\,
\left(\begin{array}{c} \ell_p\\[.1cm]
\ell_q\end{array}\right), \qquad
X=
\left(\begin{array}{cc} A(\alg)& B(\alg)\\[.3cm]
C(\alg)&D(\alg)\end{array}\right)
=
\left(\begin{array}{cc} \frac{\sin\gamma}{\sin\delta} &
\frac{\sin(\delta+\beta)}{\sin\delta}\\[.3cm]
\frac{\sin(\delta+\gamma)}{\sin\delta} &
\frac{\sin\beta}{\sin\delta}\end{array}\right) \;
\label{lin-pr}
\end{equation}
where $\alg=\{\alpha,\beta,\gamma,\delta\}$ denotes the set of angles
and $\ell_p,\ell_q,\ell_{p'},\ell_{q'}$  denote the edge lengths,
arranged as in Fig.\ref{fig-kvadrat2}. Note that due to
\eqref{2pi} the entries of the two by two matrix in \eqref{lin-pr}
satisfy the relation
\begin{equation}
AD-BC=(AB-CD)/(DB-AC).\label{constraint}
\end{equation}

Assume that the lengths $\ell_p, \ell_q$, $\ell_r$, on one
side of the two pictures in Fig.\ref{fig-cube3} are given. Let us
find the other three lengths $\ell_{p'}, \ell_{q'}$,
$\ell_{r'}$ on their opposite side, by iterating the relation
\eqref{lin-pr}. Obviously, this can be done in two different ways:
either using the front three faces, or the back ones --- the results
must be the same. This is exactly where the geometry gets into play. The
results must be consistent due to the very existence of the cube
in Fig.~\ref{fig-cube1} as a geometric body.
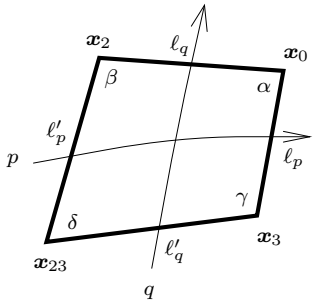
\begin{figure}[ht]
\centering
\setlength{\unitlength}{0.35mm}
\begin{picture}(100,108)
\put(0,10){\begin{picture}(100,100)
\allinethickness{.6mm}
 \path(0,10)(80,20)(90,75)(20,80)(0,10)
 \put(-5,0){\scriptsize $\boldsymbol{x}_{23}$}
 \put(80,10){\scriptsize $\boldsymbol{x}_3$}
 \put(15,85){\scriptsize $\boldsymbol{x}_2$}
 \put(90,80){\scriptsize $\boldsymbol{x}_0$}
 \put(8,15){\scriptsize $\delta$}
 \put(72,25){\scriptsize $\gamma$}
 \put(22,70){\scriptsize $\beta$}
 \put(80,67){\scriptsize $\alpha$}
 \thinlines
 \spline(-5,40)(50,50)(100,50)\path(90,52)(100,50)(90,48)
 \put(-15,40){\scriptsize $p$}
 \spline(40,0)(50,55)(60,100)\path(55,92)(60,100)(61,90)
 \put(37,-10){\scriptsize $q$}
 \put(45,5){\scriptsize $\ell_q'$}
 \put(47,83){\scriptsize $\ell_q$}
 \put(0,50){\scriptsize $\ell_p'$}
 \put(90,40){\scriptsize $\ell_p$}
\end{picture}}
\end{picture}
\caption{The angles $\alg=\{\alpha,\beta,\gamma,\delta\}$ and sides
$\ell_p,\ell_q,\ell_{p'},\ell_{q'}$ of a quadrilateral and the oriented
  rapidity lines.}
\label{fig-kvadrat2}
\end{figure}
However, they will be consistent only if all geometric relations
between the two sets of angles in the front and back faces of the
cube are taken into account. To write these relations in a
convenient form we need to introduce additional notations. Note,
that Fig.\ref{fig-kvadrat2} shows two thin lines, labeled by the
symbols ``$p$'' and ``$q$''. Each line crosses a pairs of opposite
edges, which we call ``corresponding'' (in the sense that they
correspond to the same thin line). Eq.\eqref{lin-pr} relates the
lengths $(\ell_p,\ell_q)$ of two adjacent edges with the
corresponding lengths $(\ell'_p,\ell'_q)$ on the opposite side of
the quadrilateral.

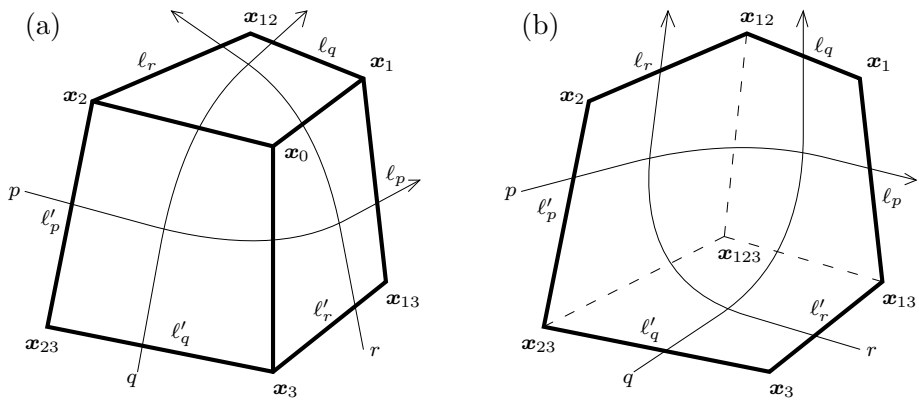
\begin{figure}[ht]
\centering
\setlength{\unitlength}{0.3mm}
\begin{picture}(400,170)
\put(230,10){\begin{picture}(150,150)
\allinethickness{.6mm}
 \path(0,20)(100,0)(150,40)(140,130)(90,150)(20,120)(0,20)
\thinlines
 \dashline{5}(0,20)(80,60)
 \dashline{5}(150,40)(80,60)
 \dashline{5}(90,150)(80,60)
 \put(-10,10){\scriptsize $\boldsymbol{x}_{23}$}
 \put(100,-10){\scriptsize $\boldsymbol{x}_{3}$}
 \put(150,30){\scriptsize $\boldsymbol{x}_{13}$}
 \put(143,135){\scriptsize $\boldsymbol{x}_{1}$}
 \put(87,155){\scriptsize $\boldsymbol{x}_{12}$}
 \put(7,120){\scriptsize $\boldsymbol{x}_{2}$}
 \put(77,50){\scriptsize $\boldsymbol{x}_{123}$}
%
 \spline(-10,80)(85,105)(165,85)\path(160,88)(165,85)(160,83)
 \spline(40,0)(115,50)(115,160)\path(112,153)(115,160)(117,153)
 \spline(140,10)(40,40)(55,160)\path(52,153)(55,160)(58,153)
 \put(-17,78){\scriptsize $p$}
 \put(35,-5){\scriptsize $q$}
 \put(143,7){\scriptsize $r$}
 \put(150,75){\scriptsize $\ell_p$}
 \put(120,142){\scriptsize $\ell_q$}
 \put(40,135){\scriptsize $\ell_r$}
 \put(118,25){\scriptsize $\ell_r'$}
 \put(43,17){\scriptsize $\ell_q'$}
 \put(-3,70){\scriptsize $\ell_p'$}
 \put(-10,150){(b)}
\end{picture}}
\put(10,10){\begin{picture}(150,150)
\allinethickness{.6mm}
 \path(0,20)(100,0)(150,40)(140,130)(90,150)(20,120)(0,20)
 \path(20,120)(100,100)
 \path(100,0)(100,100)
 \path(140,130)(100,100)
\thinlines
 \put(-10,10){\scriptsize $\boldsymbol{x}_{23}$}
 \put(100,-10){\scriptsize $\boldsymbol{x}_{3}$}
 \put(150,30){\scriptsize $\boldsymbol{x}_{13}$}
 \put(143,135){\scriptsize $\boldsymbol{x}_{1}$}
 \put(87,155){\scriptsize $\boldsymbol{x}_{12}$}
 \put(7,120){\scriptsize $\boldsymbol{x}_{2}$}
 \put(105,95){\scriptsize $\boldsymbol{x}_{0}$}
 \spline(-10,80)(100,50)(165,85)\path(160,85)(165,85)(160,80)
 \spline(40,0)(60,110)(115,160)\path(110,159)(115,160)(113,153)
 \spline(140,10)(120,115)(55,160)\path(58,155)(55,160)(60,159)
 \put(-17,78){\scriptsize $p$}
 \put(35,-5){\scriptsize $q$}
 \put(143,7){\scriptsize $r$}
 \put(150,85){\scriptsize $\ell_p$}
 \put(120,142){\scriptsize $\ell_q$}
 \put(40,135){\scriptsize $\ell_r$}
 \put(118,25){\scriptsize $\ell_r'$}
 \put(55,15){\scriptsize $\ell_q'$}
 \put(-3,65){\scriptsize $\ell_p'$}
 \put(-10,150){(a)}
\end{picture}}
\end{picture}
\caption{The ``front'' (a) and ``back'' (b) faces of the cube in
Fig.~\ref{fig-cube1} and ``rapidity'' lines.}\label{fig-cube3}
\end{figure}

Consider now Fig.\ref{fig-cube3}a which contains three directed thin lines
connecting corresponding edges of the three quadrilateral faces.
By the analogy with
the 2D Yang-Baxter equation, where similar arrangements
occur, we call them ``rapidity''
lines\footnote{%
However, at the moment we do not assume any further meaning for
these lines apart from using them as a convenient way of labeling
to the corresponding (opposite) edges of quadrilaterals.}. We will
now apply \eqref{lin-pr} three times starting from the top face
and moving against the directions of the arrows.  Introduce the
following three by three matrices
\begin{equation}
X_{pq}(\alg)=\left(\begin{array}{ccc} A& B
& 0 \\ C& D& 0 \\
0 & 0 & 1\end{array}\right)\;,\quad
X_{pr}(\alg)=\left(\begin{array}{ccc} A & 0 &
    B
\\ 0 & 1 & 0
\\ C& 0 &  D\end{array}\right)\;,
\quad
X_{qr}(\alg)=\left(\begin{array}{ccc} 1 & 0 & 0 \\ 0 & A& B \\
0 & C& D\end{array}\right)\;,
\end{equation}
where $A,B,C,D$ are defined in \eqref{lin-pr} and their dependence
on the angles $\alg=\{\alpha,\beta,\gamma,\delta\}$ is implicitly
understood. It follows that
\begin{equation}
  (\ell'_p,\ell'_q,\ell'_r)^t\,=\,X_{pq}(\alg_1)\,X_{pr}(\alg_2)
  \,X_{qr}(\alg_3)\ (\ell_p,\ell_q,\ell_r)^t
\end{equation} where
\begin{equation}
\alg_j=\{\alpha_j,\beta_j,\gamma_j,\delta_j\},\quad j=1,2,3,
\label{A-def}
\end{equation}
the lengths $\ell_p,\ell_q,\ldots$ are defined as in
Fig.\ref{fig-cube3}, and the superscript ``$t$'' denotes the
matrix transposition. Performing  similar calculations for the
back faces in Fig.\ref{fig-cube3}b and equating the resulting
three by three matrices, one obtains
\begin{equation}
X_{pq}(\alg_1)\,X_{pr}(\alg_2) \,X_{qr}(\alg_3)\,=\,
X_{qr}(\alg'_3)\, X_{pr}(\alg'_2)\, X_{pq}(\alg'_1)\ .\label{lybe}
\end{equation}
where
\begin{equation}
\alg'_j=\{\alpha'_j,\beta'_j,\gamma'_j,\delta'_j\},\quad j=1,2,3\ .
\label{Ap-def}
\end{equation}
This matrix relation contains exactly nine scalar equations where the 
LHS only depends on the front angles \eqref{A-def}, while the RHS only
depends on the back angles \eqref{Ap-def}. Solving these equations
one can obtain explicit form of the map \eqref{map}. The resulting
expressions are rather complicated and not particularly useful.
However the mere fact that the map \eqref{map} satisfy
a very special Eq.\eqref{lybe} is extremely important. Indeed, rewrite this
equation as
\begin{equation}
X_{pq}(\alg_1)\,X_{pr}(\alg_2) \,X_{qr}(\alg_3)\,=\,\R_{123}\,
\Big( X_{qr}(\alg_3)\, X_{pr}(\alg_2)\, X_{pq}(\alg_1)\Big)
\label{tza}
\end{equation}
where $\R_{123}$ is an operator acting as the  substitution
\eqref{map} for any function
$F(\alg_1,\alg_2,\alg_3)$ of the angles,
\begin{equation}
\R_{123}\Big(F(\alg_1,\alg_2,\alg_3)\Big)=F(\alg_1',\alg_2',\alg_3')
\end{equation}
Then, following the arguments of \cite{Korepanov:1993jsp}, one can
show that the map \eqref{map} satisfies the {\em functional
tetrahedron equation} \cite{KashaevKorepanovSergeev}
\begin{equation}
\R_{123} \cdot\R_{145}\cdot \R_{246} \cdot\R_{356} \;=\;
\R_{356}\cdot \R_{246} \cdot\R_{145} \cdot\R_{123}\label{fte}\ ,
\end{equation}
where both sides are
compositions of the maps \eqref{map}, involving six different sets
of angles. Algebraically, this equation arises
as an associativity condition for the
cubic algebra \eqref{tza}.
To discuss its geometric meaning
we need to introduce {\em discrete evolution systems} associated
with the map \eqref{map}.

\subsection{Discrete evolution systems: ``Existence as
  integrability''}

\begin{figure}[htb]
\centering
\includegraphics[scale=0.55]{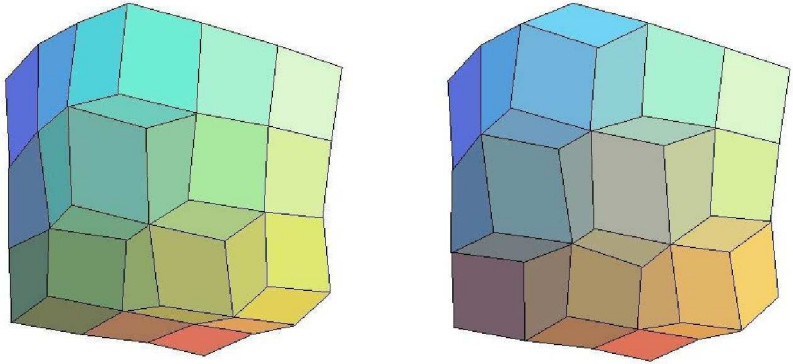}
\caption{Visualization of the 3D ``staircase'' evolution.}\label{fig-viz}
\end{figure}

Consider a sub-lattice $L$ of the 3D quadrilateral lattice,
which only includes points $\x(m)$ with
$m_1,m_2,m_3\ge0$.  The boundary of this sub-lattice is a 2D discrete
surface formed by quadrilaterals with the vertices
$\x(m)$ having at least one of their integer
coordinates $m_1,m_2,m_3$ equal to zero and the other two
non-negative.
Assume that all quadrilateral angles on this surface are known, and
consider them as initial data. Then repeatedly applying the
map \eqref{map} one can calculate angles on all faces of the
sub-lattice $L$, defined above (one has to start from the corner
${\x}(0)$).
The process can be visualized as an evolution of
the initial data surface where
every transformation \eqref{map} corresponds to
a ``flip'' between the front and back faces (Fig.~\ref{fig-cube2})
of some cube adjacent to the surface.
This makes the
surface looking as a 3D ``staircase'' (or a pile of cubes) in the
intersection corner of the three coordinate planes,
see Fig.~\ref{fig-viz} showing  two stages of this process.
Note, that the corresponding evolution equations can be written in a
covariant form for an arbitrary lattice cube (see Eq.\eqref{covariant}
below for an example). It is also useful to have in mind that the
above evolution can be defined purely geometrically as a
{\em ruler-and-compass\/} type construction. Indeed the construction of the
point ${\x}_{123}$ in Fig.~\ref{fig-cube1} from the points
$\x_0,\x_1,\x_2,\x_3,\x_{12},\x_{13},\x_{23}$
(and that is what is necessary for flipping a cube)
only requires a {\em 2D-ruler\/} which allows to draw planes through
any three non-collinear points in the Euclidean space.

\begin{figure}[ht]
\centering
\includegraphics[scale=0.60]{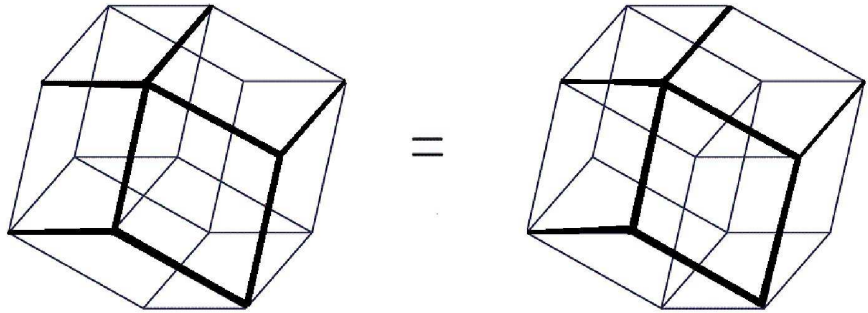}
\caption{Two dissections of the rhombic dodecahedron into four
quadrilateral hexahedra.} \label{fig-tetra}
\end{figure}

Similar evolution systems can be defined for other
quadrilateral lattices instead
of the 3D cubic lattice considered above.
Since the evolution is local (only one cube is flipped at a
time) one could consider finite lattices as well. For example, consider
six adjacent quadrilateral faces covering the front surface of the
rhombic dodecahedron\footnote{%
It is worth noting that the most general rhombic dodecahedron
with quadrilateral faces can only be embedded into (at least)
the 4D Euclidean space.}
shown in Fig.~\ref{fig-tetra}. Suppose that all  angles on
these faces are given and consider them as initial data.
Now apply a sequence of four maps \eqref{map} and calculate angles on
the back surface of the rhombic dodecahedron.
This can be done in two alternative ways, corresponding to the two
different dissection of the rhombic dodecahedron into four cubes shown
in Fig.\ref{fig-tetra}.
The functional tetrahedron equation \eqref{fte} states
that the results will be the same. Thereby it gives an algebraic proof for
the equivalence of two ``ruler-and-compass'' type constructions of the
back surface of the dodecahedron in Fig.~\ref{fig-tetra}.
Can we also prove this equivalence geometrically?
Although from the first sight this
does not look trivial, it could be easily done from the point of view of
the 4D geometry.
The required statement follows just from the fact of existence of the
quadrilateral lattice with the topology of the 4D cube
\cite{DoliwaSantini}. The latter is
defined by eight intersecting 3-planes in a general position in the 4-space.
The two rhombic dodecahedra shown in Fig.~\ref{fig-tetra}
are obtained by a dissection of the 3-surface of this 4-cube, along its
2-faces, so these dodecahedra
must have exactly the same quadrilateral 2-surface.
Thus the functional tetrahedron equation \eqref{fte}, which plays the
role of integrability condition for the discrete evolution system
associated with the map \eqref{map}, simply follows from the mere fact of
existence of the 4-cube, which is the simplest 4D quadrilateral lattice.
For a further discussion of a relationship between the geometric
consistency and integrability see \cite{BoSurbook}.

\section{Quantization of the 3D circular
 lattices}
\subsection{Poisson structure of circular lattices}\label{poisson-sec}
\begin{figure}[hbt]
\centering
\includegraphics[scale=.61]{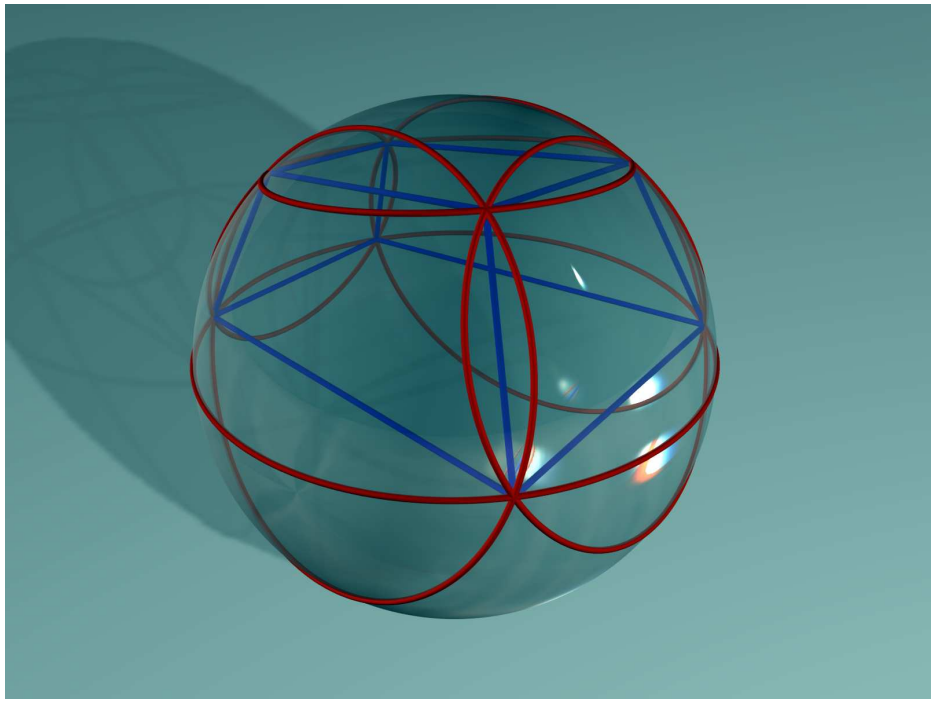}
\caption{Miquel configuration of circles in 3D space, an elementary
hexahedron and its circumsphere.}\label{miguel}
\end{figure}

The 3D {\em circular lattice} \cite{Bob96,CDS97,KS98} is a
special 3D quadrilateral lattice where all faces are circular
quadrilaterals (i.e., quadrilaterals which can be inscribed into a
circle).  The existence of these lattices is established by the
following beautiful geometry theorem due to Miquel \cite{Miquel}
(see Fig.~\ref{miguel})

\begin{quote}{\bf Miquel theorem.} {\em Consider four
points $\x_0, \x_1, \x_2, \x_3$ in general position in ${\mathbb
R}^N$, $N\ge 3$. On each of the three circles $c(\x_0,\x_i,\x_j)$,
$1\le i <j\le 3$ choose an additional new point $\x_{ij}$. Then
there exist a unique point $\x_{123}$ which simultaneously belongs
to the three circles $c(\x_1,\x_{12},\x_{13})$,
$c(\x_2,\x_{12},\x_{23})$ and $c(\x_3,\x_{13},\x_{23})$.}
\end{quote}
It is easy to see that the above six circles lie on the same
sphere. It follows then that every elementary
``cube'' on a circular lattice (whose vertices are
at the circle intersection points) is inscribable into a sphere, see
Fig.~\ref{miguel}.
The general formulae of the previous subsection can be readily
specialized for the circular lattices. A circular quadrilateral has
only two independent angles.
In the notation of Fig.~\ref{fig-kvadrat2} one has
\begin{equation}
\gamma=\pi-\beta,\qquad \delta=\pi-\alpha\ .\label{circ-ang}
\end{equation}
Due to the Miquel theorem we can simply impose these restrictions on
all faces of the lattice without running to any contradictions.
The two by two matrix in \eqref{lin-pr} takes the form
\begin{equation}
X\;=\;\left(\begin{array}{cc} k & a^* \\[.2cm]
                             -a & k
\end{array}\right),
\qquad \det \, X=1\label{detx}\ ,
\end{equation}
where we have introduced new variables
\begin{equation}
k\;=\;(\csc \alpha)\, {\sin\beta}\;,\quad
a=({\csc\alpha})\,{\sin(\alpha+\beta)}\;,\quad
a^*=({\csc\alpha})\,{\sin(\alpha-\beta)},\label{newvar}
\end{equation}
instead of the two angles $\{\alpha,\beta\}$.
Note that the new variables  are constrained by the relation
\be
aa^*=1-k^2\ . \label{aak-rel}
\ee
Conversely, one has
\begin{equation}
\cos\alpha=\frac{a-a^*}{2k}\;,\quad \cos\beta=\frac{a+a^*}{2}\ .
\end{equation}
Let the variables $\{k_j,a_j,a_j^*\}$,
$\{k'_j,a'_j,a^{*\prime}_j\}$, $j=1,2,3$,\   correspond to the front
and back faces of the cube. The map \eqref{map} then read
explicitly
\begin{equation}\label{themap}
{\mathcal R}_{123}:\qquad
\left\{\begin{array}{ll}
(k_2^{}a_1^*)'=\; k_3^{}a_1^* - \varepsilon k_1^{}a_2^*a_3^{}, &
(k_2^{}a_1^{})'=k_3^{}a_1^{}- \varepsilon k_1^{}a_2^{}a_3^*, \\\\
(a_2^*)'  = a_1^*a_3^* + \varepsilon k_1^{}k_3^{}a_2^*, &
(a_2^{})' = a_1^{}a_3^{} + \varepsilon k_1^{}k_3^{}a_2^{}, \\\\
(k_2^{}a_3^*)'  = k_1^{}a_3^* - \varepsilon k_3^{}a_1^{}a_2^*, &
(k_2^{}a_3^{})' = k_1^{}a_3^{} - \varepsilon k_3^{}a_1^*a_2^{},
\end{array}\right.
\end{equation}
where $\varepsilon=+1$ and
\be
k_2^{\prime}=\sqrt{1-a_2'a_2^{*\prime}}%
\ .
\label{aak2}
\ee

At this point we
note that exactly the same map together with the corresponding equations
\eqref{lybe} and \eqref{fte} were previously obtained in \cite{BS05}.
Moreover, it was discovered that this map is a canonical transformation
preserving the Poisson algebra
\begin{equation}\label{o-poisson}
\{a_i,a_j^*\}\;=\;2\,\delta_{ij}\,k_i^2\;, \quad \{k_i,a_j\}=
\delta_{ij}\,k_i\,a_i\;, \quad
\{k_i,a_j^*\}=-\,\delta_{ij}\,k_i\,a_i^*,\qquad i,j=1,2,3\ ,
\end{equation}
where $k_i^2=1-a_ia_i^*$. Note that variables $k,a,a^*$ on different
quadrilaterals are in involution.
The same Poisson algebra in terms of angle variables reads
\begin{equation}\label{poisson}
\{\alpha_i,\beta_j\}= \delta_{ij}\;,\quad
\{\alpha_i,\alpha_j\}=\{\beta_i,\beta_j\}=0\;.
\end{equation}
This  ``ultra-local'' symplectic structure trivially extends to
any circular quad-surface of initial data, discussed above.
To resolve an apparent ambiguity in naming of the angles,
this surface must be equipped with oriented rapidity lines, similar to
those in Fig.~\ref{fig-cube3}\footnote{We refer
  the reader to our previous paper \cite{BMS07a} where the relationship
  between the rapidity graphs and quadrilateral lattices is thoroughly
  discussed.}.
In addition, the angles for each quadrilateral should be arranged as in
Fig.~\ref{fig-kvadrat2}.
Then one can assume
that the indices $i,j\/$\ \ in \eqref{poisson}   refer to all
quadrilaterals on this surface.

Thus,
the evolution defined by the map \eqref{themap} is a symplectic
transformation. The corresponding equations of motion for the whole
lattice (the analog of the Hamilton-Jacobi equations) can be written
in a ``covariant'' form. For every cube define
\begin{equation}
A_{32}^{}=a_1^{},\;\; A_{23}^{}=a_1^*,\;\;
A_{31}^{}=a_2',\;\; A_{13}^{}=a_2^{'*},\;\;
A_{21}^{}=a_3^{},\;\; A_{12}^{}=a_3^*\label{cov-form}
\end{equation}
where $A_{jk}$ stands for $A_{jk}({m})$, where $m$ is such that ${\x}({m})$
coincides with the coordinates of the top front corner of the cube
(vertex ${\x}_0$ in Fig.\ref{fig-cube1}).
Let $T_k$ be the shift operator $T_k\, A_{ij}({m})=A_{ij}({m}+{e}_k)$. Then
\begin{equation}\label{covariant}
\widetilde{T}_k A_{ij} \;=\;
\frac{A_{ij}-A_{ik}A_{kj}}{K_{ik}K_{kj}}\;,\quad
K_{ij}=K_{ji}=\sqrt{1-A_{ij}A_{ji}}\;,
\end{equation}
where $(i,j,k)$ is an arbitrary permutation of $(1,2,3)$ and
\begin{equation}
\widetilde{T}_1^{}=T_1^{}\;,\quad
\widetilde{T}_2^{}=T_2^{-1}\;,\quad \widetilde{T}_3^{}=T_3^{}\ .
\end{equation}
Note that Eq.(\ref{covariant}) also imply
\be
(\widetilde{T}_k K_{ij})K_{kj} = (\widetilde{T}_i K_{kj}) K_{ij}\ .
\label{kk-rel}
\ee

{\bf Remarks}. The equations \eqref{covariant} have been previously obtained 
in \cite{KS98}, see Eq.(7.20) therein. The quantities $A_{ij}$ in
\eqref{covariant} should be identified with the {\em rotation
  coefficients} denoted as $\tilde\beta_{ij}$ in \cite{KS98}.
The same equations \eqref{covariant} are discussed in \S3.1 of
\cite{BoSurbook}, where one can also find a detailed bibliography on the
circular lattices (we are indebted to A.I.Bobenko for these
important remarks).

\subsection{Quantization and the tetrahedron equation}\label{quantization}
In the next section we consider different quantizations of the map
\eqref{themap} and obtain several solutions of the full quantum
tetrahedron equation (see Eq.\eqref{TE} below).
In all cases we start with the canonical quantization of the Poisson
algebra \eqref{poisson},
\be
[\alpha_i,\beta_j]=\xi\, \hbar \,\delta_{ij}\,,\qquad
[\alpha_i,\alpha_j]=0\,,\qquad
[\beta_i,\beta_j]=0\,,\label{ang-alg}
\ee
where $\hbar$ is the quantum parameter (the Planck constant) and $\xi$
is a numerical coefficient, introduced for a further convenience. The
indices $i,j$ label the faces of the ``surface of initial data''
discussed above.
Since the commutation relations \eqref{ang-alg} are ultra-local (in
the sense that the angle variables on different faces commute with
each other), let us concentrate on the local Heisenberg algebra,
\be
{\mathsf H}:\qquad\qquad  [\alpha,\beta]=\xi\, \hbar\ ,\label{H-sing}
\ee
for a single lattice face (remind that the angles
shown in Fig.~\ref{fig-kvadrat2} are related by \eqref{circ-ang}).
The map \eqref{themap} contains the quantities $k,a,a^*$,
defined in \eqref{newvar}, which
now become operators. For definiteness, assume that the
non-commuting factors in \eqref{newvar} are ordered exactly as
written. Then the definitions \eqref{newvar} give
\begin{eqnarray}
k\phantom{{}^*}&=&(U-U^{-1})^{-1}\, (V-V^{-1}),\nonumber\\[0.2cm]
a\phantom{{}^*}&=&q^{-\frac{1}{2}}\,(U-U^{-1})^{-1}\,
(U\,V-U^{-1}\,V^{-1}),\label{aa-w}
\\[0.2cm]
a^*&=&q^{+\frac{1}{2}}\,(U-U^{-1})^{-1}\, (U\,V^{-1}-U^{-1}\,V),\nonumber
\end{eqnarray}
where the elements $U$ and $V$ generate the Weyl algebra,
\be
U\,V=q\,V\,U,\qquad U=e^{i\alpha},\qquad
V=e^{i\beta},\qquad q=e^{-\xi\hbar}\ .\label{weyl1}
\ee
The operators \eqref{aa-w} obey the commutation relations of the
$q$-oscillator algebra,
\be\label{q-osc1}
\mathsf{Osc}_{\,q}:\qquad
\left\{
\renewcommand\arraystretch{2.0}
\begin{array}{l}
q\,a^* \, a- q^{-1}\, a \,a^*=q-q^{-1},
\qquad k\,a^*=q\,a^*\,k,\qquad k\,a=q^{-1}\,a\,k\;,\\
k^2=q\,(1-a^*\, a)=q^{-1}\,(1-a\, a^*)\ .
\end{array}\right.
\renewcommand\arraystretch{1.0}
\ee
where the element $k$ is assumed to be invertible.
This algebra is, obviously,
a quantum counterpart of the
Poisson algebra (\ref{o-poisson}).
In the previous Section we have already mentioned the result of
\cite{BS05} that
\begin{itemize}
\item[(i)] the map \eqref{themap} is
an automorphism of the tensor cube of the Poisson algebra
\eqref{o-poisson} (remind that the relation \eqref{aak-rel} should be
taken into account in \eqref{themap}).
\end{itemize}
In the same paper \cite{BS05} it was also shown that
\begin{itemize}
\item[(ii)]
there exists a quantum version of the map \eqref{themap}, which
acts as an automorphism of the tensor cube of
the $q$-oscillator algebra \eqref{q-osc1}.
The formulae \eqref{themap} for the quantum map stay exactly the same, but
the relation
\eqref{aak-rel} should be replaced by either of the two relations on
the second line of \eqref{q-osc1}, for instance,
$k^2=q\,(1-a^* a)$. In particular, \eqref{aak2} should be replaced with
\be
(k_2')^2=q\,(1-{a_2^*}^{\prime} {a_2}')\ .\label{aak2q}
\ee
\item[(iii)]
the quantum version of the map \eqref{themap}, defined in (ii) above, satisfies
the functional tetrahedron equation \eqref{fte}. 
\end{itemize}

For any irreducible representations of the $q$-oscillator algebra
\eqref{q-osc1} the formulae \eqref{themap} and \eqref{aak2q}
uniquely determine the
$\R_{123}$ as
an internal automorphism,
\begin{equation}\label{rl-def}
\R_{123}\big( F \big)\;=\; R_{123}^{} \;F\; R_{123}^{-1},\qquad
 F\in \mathsf{Osc}_q\otimes\mathsf{Osc}_q\otimes\mathsf{Osc}_q\ .
\end{equation}
It follows then from \eqref{fte} that the linear operator $R_{123}$
 satisfies the quantum tetrahedron equation
\begin{equation}\label{TE}
R_{123}\;R_{145}\;R_{246}\;R_{356}\;=\;
R_{356}\;R_{246}\;R_{145}\;R_{123}\;,
\end{equation}
where each of the operators $R_{123}$, $R_{145}$, $R_{246}$ and $R_{356}$ act
as \eqref{rl-def} in the three factors (indicated by the subscripts) 
of a tensor product
of six $q$-oscillator algebras and
act as the unit operator in the remaining three factors.

\section{Solutions of the tetrahedron equation}
\label{solutions}
Here we show that all known solutions of the tetrahedron equation can
be obtained by solving \eqref{rl-def} for the linear operator
$R_{123}$ with different irreducible
representations of the $q$-oscillator algebra \eqref{q-osc1}.

\subsection{Fock representation solution}

In this subsection we set $\xi=1$ in \eqref{ang-alg} and $q=e^{-\hbar}$. 
Define the Fock representation of a single $q$-oscillator
algebra \eqref{q-osc1}, 
\begin{equation}
a\,|0\rangle=0,\quad
a\,|n+1\rangle =|n\rangle,\quad a^*\,|n\rangle = (1-q^{2+2n})\,
|n+1\rangle,\quad k\,|n\rangle = q^{n+1/2}\,|n\rangle,
\end{equation}
spanned by the vectors $|n\rangle$, \ $n=0,1,\ldots,\infty$.
Then using \eqref{themap}, \eqref{q-osc1}, \eqref{aak2q}
and \eqref{rl-def} one can show that 
\begin{equation}\label{R-fock}
\renewcommand\arraystretch{2.2}
\begin{array}{ll}
\ds \langle n_1,n_2,n_3|R\,|n'_1,&n'_2,n_3\rangle =
\ds \delta_{n_1^{}+n_2^{},n_1'+n_2'}
\delta_{n_2^{}+n_3^{},n_2'+n_3'}\;\,(-1)^{n_2}\,q^{(n_1'-n_2^{})
(n_3'-n_2^{})}\\
&\ds\times \;
\binom{n_3}{n_2'}_{\!q^2}
\,{}_2\phi_1(q^{-2n_2'},q^{2(1+n_3')},q^{2(1-n_2'+n_3)};q^2,q^{2(1+n_1)})\ ,
\end{array}
\renewcommand\arraystretch{1.0}
\end{equation}
where
\begin{equation}
(x;q^2)_n\;=\;\prod_{j=0}^{n-1}(1-q^{2j}x)\;,\quad
\binom{n}{k}_{\!q^2}=\frac{(q^2;q^2)_n}{(q^2;q^2)_k(q^2;q^2)_{n-k}}\;,
\end{equation}
and
\begin{equation}\label{q-gauss}
\phantom{f}_2\phi_1(a,b,c;q^2,z)\;=\;
\sum_{n=0}^{\infty}
\frac{(a;q^2)_n(b;q^2)_n}{(q^2;q^2)_n(c;q^2)_n} z^n
\end{equation}
is the $q$-deformed Gauss hypergeometric series.
This 3D $R$-matrix satisfies the constant tetrahedron equation
(\ref{TE}). In matrix form this equation reads 
\begin{equation}
\label{te-fock}
\begin{array}{ll}
\ds\sum_{n_j'=0}^\infty R_{n_1^{},n_2^{},n_3^{}}^{n_1',n_2',n_3'}\,
&R_{n_1',n_4^{},n_5^{}}^{n_1'',n_4',n_5'}\,
R_{n_2',n_4',n_6^{}}^{n_2'',n_4'',n_6'}\,
R_{n_3',n_5',n_6'}^{n_3'',n_5'',n_6''} =\\
&= \ds\sum_{n_j'=0}^\infty
R_{n_3^{},n_5^{},n_6^{}}^{n_3',n_5',n_6'}\,
R_{n_2^{},n_4^{},n_6'}^{n_2',n_4',n_6''}\,
R_{n_1^{},n_4',n_5'}^{n_1',n_4'',n_5''}\,
R_{n_1',n_2',n_3'}^{n_1'',n_2'',n_3''}
\end{array}
\end{equation}
where the sum is taken over six indices
$n'_1,n'_2,n'_3,n'_4,n'_5,n'_6$ and
\begin{equation}
R_{n_1,n_2,n_3}^{n_1',n_2',n_3'}=\langle n_1,n_2,n_3|R\,|n_1',n_2',n_3'\rangle
\ .
\end{equation}
Note that Eq.\eqref{te-fock} does not contain any spectral parameters.
Originally the $R$-matrix \eqref{R-fock} was obtained in
\cite{BS05} in terms of a solution of some recurrence relation, which 
was subsequently reduced to the
$q$-hypergemetric function in \cite{BMS08a}.

\subsection{Modular double solution}\label{modular-sec}

In this subsection we set in \eqref{ang-alg}
\be
\xi=-i,\qquad \hbar=\pi\,b^2,\qquad q=e^{i\pi b^2}
\ee
where $b$ is a free parameter, $\mathrm{Re}\, b\not=0$. Here it will
be more
convenient to work with a slightly modified version\footnote{%
Note that Eq.\eqref{themap} is a particular
case a general map considered in \cite{BS05}; see also 
Eq.\eqref{TE-aux} below.}
of the map \eqref{themap}, with the value $\varepsilon =-1$.

Consider a non-compact representation \cite{Schmud94} of the $q$-oscillator
algebra \eqref{q-osc1} in the space of functions $f(\sigma)\in {\mathbb
  L}^2({\mathbb R})$ on the real line admitting an analytical
continuation into an appropriate horizontal
strip, containing the real axis in the
complex $\sigma$-plane (see
\cite{Schmud94} for further details).
Such representation essentially reduces to that of the Weyl algebra
\be
{\mathsf W}_q:\qquad\qquad k\,w =q\, w\, k,\qquad q=e^{i\pi b^2},\label{weyl2}
\ee
realized as multiplication and shift operators
\be
k\,|\sigma\rangle = \ii
e^{\pi\sigma b}\, |\sigma\rangle\;,\qquad w\,
|\sigma\rangle = |\sigma-\ii b\rangle. \label{w-rep}
\ee
The generators $a,a^*$ in \eqref{q-osc1} are expressed as
\be
a=w^{-1}\;,\quad a^*=(1-q^{-1}k^2)\,w\;.
\ee
As explained in \cite{Faddeev:1995} the representation \eqref{w-rep}
is not, in general, irreducible. Therefore, the relation
\eqref{rl-def} alone does not unambiguously define 
the linear operator $R_{123}$ in this case. 
Following the idea of \cite{Faddeev:1995} consider the modular 
dual of the algebra \eqref{weyl2}, 
\be
{\mathsf W}_{\tilde{q}}:\qquad\qquad \tilde{k}\,\tilde{w} =\tilde{q}\,
\tilde{w}\, \tilde{k},\qquad \tilde{q}=e^{-i\pi b^{-2}},\label{weyl3}
\ee
acting in the same representation space
\be
\tilde{k}\,|\sigma\rangle = -\ii
e^{\pi\sigma b^{-1}}\, |\sigma\rangle\;,\qquad \tilde{w}\,
|\sigma\rangle = |\sigma+\ii b^{-1}\rangle.
\ee
We found that if the 
relation \eqref{rl-def} is complemented by its modular dual
\begin{equation}\label{rl-def1}
\tilde{\R}_{123}\big( \tilde{F} \big)
\;=\; R_{123}^{} \;\tilde{F}\; R_{123}^{-1},\qquad
 \tilde{F}\in
 \mathsf{Osc}_{\tilde{q}}
\otimes\mathsf{Osc}_{\tilde{q}}\otimes\mathsf{Osc}_{\tilde{q}}\ ,
\end{equation}
then the pair of relations \eqref{rl-def} and \eqref{rl-def1}
determine the operator $R_{123}$ uniquely\footnote{It is worth
  mentioning similar phenomena in
  the construction of the $R$-matrix \cite{Faddeev:1999} for the
  modular double of the quantum group $U_q(sl_2)$ and the
  representation theory of $U_q(sl_2,{\mathbb R})$ \cite{PT99}.}.
The dual $q$-oscillator
algebra $\mathsf{Osc}_{\tilde{q}}$
is realized through the dual Weyl pair \eqref{weyl3} and the relations
\be
\tilde{a}=\tilde{w}^{-1}\;,\quad
\tilde{a}^*=(1-\tilde{q}^{-1}\tilde{k}^2)\,\tilde{w}\;.
\end{equation}
The dual version of the map $\tilde{\R}_{123}$ is defined by the same
formulae \eqref{themap}, where quantities
$k_j,a_j,a^*_j$, \ $j=1,2,3$ are replaced by their ``tilded''
counterparts $\tilde{k}_j,\tilde{a}_j,\tilde{a}^*_j$. The value of $q$
does not, actually, enter the map \eqref{themap}, but needs to be taken
into account in the relations between the generators of the
$q$-oscillator algebra.
Thus, the linear operator $R_{123}$ in this case simultaneously provides
the two maps ${\R}_{123}$ and $\tilde{\R}_{123}$ (given by
\eqref{themap} with $\varepsilon=-1$). 
The explicit form of this operator is given below.

Denote
\be
\eta=\frac{b+b^{-1}}{2},
\ee
and define a special function
\begin{equation}\label{spec-f}
\Bigpsi{c_1,c_2}{c_3,c_4}{c_0}\;=\; \int_{\mathbb{R}} \, dz\,
e^{2\pi\ii z (-c_0-\ii\eta)}\,
\frac{\varphi(z+\frac{c_1+\ii\eta}{2})\varphi(z+\frac{c_2+\ii\eta}{2})}
{\varphi(z+\frac{c_3-\ii\eta}{2})\varphi(z+\frac{c_4-\ii\eta}{2})}\;,
\end{equation}
where $\varphi$ is the non-compact quantum dilogarithm
\cite{Faddeev:1994}
\begin{equation}
\varphi(z)=\exp\left(\ds \frac{1}{4}\int_{\mathbb{R}+\ii 0}
\frac{e^{-2\ii zx}}{\textrm{sinh}(xb)\textrm{sinh}(x/b)}\
\frac{dx}{x}\right)\;.
\end{equation}
The values of $c_1,c_2,c_3,c_4$ are assumed to be such that poles of
numerator in the integrand of \eqref{spec-f} lie above the real axis,
while the zeroes of the denominator lie below the real axis. For other
values of $c_j$ the integral \eqref{spec-f} is defined by an analytic
continuation.
Note that for $\mathrm{Im}\,b^2>0$ the integral
$\phantom{|}_2 \kern -.05em \Psi_2$ can be
evaluated by closing the integration contour in the
upper half plane (see eq.(78) in \cite{BMS08a}), which is very convenient
for numerical calculations.

The matrix elements 
of $R$-matrix solving the pair of the equations \eqref{rl-def}
and \eqref{rl-def1} are given by
\begin{equation}\label{R-modular}
\begin{array}{ll}
\langle \sigma_1,&\sigma_2,\sigma_3 
|R\,|\sigma'_1,\sigma'_2,\sigma'_3\rangle 
=\delta_{\sigma_1^{}+\sigma_2^{},\sigma_1'
+\sigma_2'}\delta_{\sigma_2^{}+\sigma_3^{},\sigma_2'+\sigma_3'}\times
\\[.3cm]
&\times\, 
\ds
e^{\ii \pi
(\sigma_1'\sigma_3'+\ii\eta(\sigma_1'+\sigma_3'-\sigma_2^{}))}
\Bigpsi{\sigma_1-\sigma_3,-\sigma_1+\sigma_3}
{\sigma_1+\sigma_3,-\sigma_1'-\sigma_3'}{\sigma_2}\ .
\end{array}
\end{equation}
This $R$-matrix satisfies the constant
tetrahedron equation (\ref{TE}). Its matrix form is given by
\eqref{te-fock} where $R_{n_i,n_j,n_k}^{n_i', n_j', n_k'}$ is
substituted by
$R_{\sigma_i,\sigma_j,\sigma_k}^{\sigma_i',\sigma_j',\sigma_k'}=
\langle
\sigma_i,\sigma_j,\sigma_k|R\,|\sigma_i',\sigma_j',\sigma_k'\rangle$,
$i,j,k=1,2,\ldots,6$, 
and the sums are replaced by the integrals over
$\sigma'_1,\sigma'_2,\sigma'_3,\sigma'_4,\sigma'_5,\sigma'_6$ 
along the real lines $-\infty<\sigma'_i<\infty$. One can verify that these 
integrals converge. 
The solution \eqref{R-modular} was obtained in \cite{BMS08a}.

\subsubsection{The ``interaction-round-a-cube'' formulation of the
  modular double solution}\label{IRC}
\begin{figure}[ht]
\centering
\setlength{\unitlength}{0.28mm}
\begin{picture}(150,150)
\put(0,10){
\begin{picture}(150,125)
\allinethickness{.6mm}
 \path(0,40)(60,0)(140,20)(120,100)(75,125)(20,110)(0,40)
 \path(60,85)(20,110)
 \path(60,85)(120,100)
 \path(60,85)(60,0)
\thinlines
 \dashline{5}(85,60)(0,40)
 \dashline{5}(85,60)(75,125)
 \dashline{5}(85,60)(140,20)
 \put(59,91){\scriptsize $\boldsymbol{a}$}
 \put(58,-10){\scriptsize $\boldsymbol{e}$}
 \put(-7,32){\scriptsize $\boldsymbol{c}$}
 \put(142,12){\scriptsize $\boldsymbol{d}$}
 \put(13,115){\scriptsize $\boldsymbol{g}$}
 \put(122,105){\scriptsize $\boldsymbol{f}$}
 \put(70,130){\scriptsize $\boldsymbol{b}$}
 \put(82,51){\scriptsize $\boldsymbol{h}$}
\end{picture}}\end{picture}
\caption{The arrangement of corner spins
around a cube.}\label{fig-cubeweight}
\end{figure}
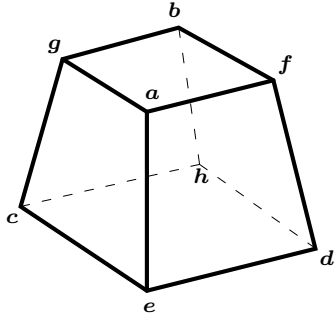

Note that due to the presence of two delta-functions in \eqref{R-modular}
the edge spins are constrained by two relations
$\sigma_1^{}+\sigma_3^{}=\sigma_1'+\sigma_3$ and 
$\sigma_2^{}+\sigma_3^{}=\sigma_2'+\sigma_3'$
at each vertex of the lattice. Here we re-formulate this solution 
in terms of unconstrained {\em corner\/}\  spins, which also
take continuous values on the real line.
Figure~\ref{fig-cubeweight} shows an elementary cube of
the lattice with the corner spins ``$a,b,c,d,e,f,g,h$'' arranged in the
same way as in \cite{Baxter:1986phd}.
The corresponding Boltzmann weight reads \cite{BMS08a}
\be\begin{array}{l}
W(a|e,f,g|b,c,d|h; \T_1,\T_2,\T_3) \;=\;
\\[.2cm]
\phantom{XXXXX}\ds=e^{\ii \pi
(\sigma_1'\sigma_3'+\ii\eta(\sigma_1'+\sigma_3'-\sigma_2^{}))}
\Bigpsi{\sigma_1-\sigma_3,-\sigma_1+\sigma_3}
{\sigma_1+\sigma_3,-\sigma_1'-\sigma_3'}{\sigma_2}\ ,\label{Wweight} 
\end{array}
\ee
where  
\begin{equation}
\begin{array}{lll}
\sigma_1=g+f-a-b-\T_1,&\sigma_2=a+c-e-g+\T_2,&
\sigma_3=e+f-a-d-\T_3,\\[.2cm]
 \sigma_1'=c+d-e-h-\T_1,&
\sigma_2'=f+h-b-d+\T_2,&\sigma_3'=b+c\/-g-h\/-\T_3
\end{array}
\end{equation}
 Note, that the variables  $\sigma_i$ and $\sigma_i'$ 
automatically satisfy the delta function constrains in \eqref{R-modular}.
The weight functions \eqref{Wweight} contains three arbitrary (spectral)
parameters
$\T_1,\T_2,\T_3$.
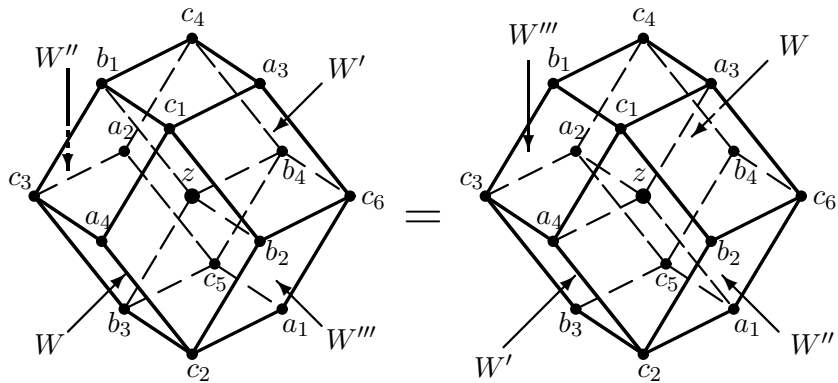
\begin{figure}[ht]
\centering
\setlength{\unitlength}{0.15mm}
\begin{picture}(700,440)
\put(0,100){\begin{picture}(300,300)
\thicklines
\multiput(141,0)(60,100){2}{\line(2,1){80}}
\multiput(140,0)(60,100){2}{\line(2,1){80}}
\multiput(139,0)(60,100){2}{\line(2,1){80}}
\multiput(140,1)(80,40){2}{\line(3,5){60}}
\multiput(140,0)(80,40){2}{\line(3,5){60}}
\multiput(140,-1)(80,40){2}{\line(3,5){60}}
\multiput(0,141)(60,-40){2}{\line(3,5){60}}
\multiput(0,140)(60,-40){2}{\line(3,5){60}}
\multiput(0,139)(60,-40){2}{\line(3,5){60}}
\multiput(1,140)(60,100){2}{\line(3,-2){60}}
\multiput(0,140)(60,100){2}{\line(3,-2){60}}
\multiput(-1,140)(60,100){2}{\line(3,-2){60}}
\multiput(0,141)(60,-40){2}{\line(4,-5){80}}
\multiput(0,140)(60,-40){2}{\line(4,-5){80}}
\multiput(0,139)(60,-40){2}{\line(4,-5){80}}
\multiput(200,101)(80,40){2}{\line(-4,5){80}}
\multiput(200,100)(80,40){2}{\line(-4,5){80}}
\multiput(200, 99)(80,40){2}{\line(-4,5){80}}
\multiput(61,240)(60,-40){2}{\line(2,1){80}}
\multiput(60,240)(60,-40){2}{\line(2,1){80}}
\multiput(59,240)(60,-40){2}{\line(2,1){80}}
\multiput(141,0)(60,240){2}{\line(-3,2){60}}
\multiput(140,0)(60,240){2}{\line(-3,2){60}}
\multiput(139,0)(60,240){2}{\line(-3,2){60}}
\multiput(160,80)(30,-20){2}{\line(3,-2){20}}
\multiput(220,180)(30,-20){2}{\line(3,-2){20}}
\multiput(80,40)(30,15){3}{\line(2,1){20}}
\multiput(0,140)(30,15){3}{\line(2,1){20}}
\multiput(160,80)(-20,25){4}{\line(-4,5){16}}
\multiput(220,180)(-20,25){4}{\line(-4,5){16}}
\multiput(220,180)(-16,-26.667){4}{\line(-3,-5){12}}
\multiput(140,280)(-16,-26.667){4}{\line(-3,-5){12}}
\multiput(140,140)(30,-20){2}{\line(3,-2){20}}
\multiput(140,140)(30,15){3}{\line(2,1){20}}
\multiput(140,140)(-20,25){4}{\line(-4,5){16}}
\multiput(140,140)(-16,-26.667){4}{\line(-3,-5){12}}
\multiput(140,0)(60,100){2}{\circle*{8}}
\multiput(220,40)(60,100){2}{\circle*{8}}
\multiput(60,100)(60,100){2}{\circle*{8}}
\multiput(0,140)(60,100){2}{\circle*{8}}
\multiput(140,280)(80,-100){2}{\circle*{8}}
\multiput(80,180)(80,-100){2}{\circle*{8}}
\multiput(80,40)(120,200){2}{\circle*{8}}
\put(140,140){\circle*{12}}
\put(255,25){\vector(-1,1){40}}\put(255,240){\vector(-1,-1){42.5}}
\put(30,20){\vector(1,1){51}}\put(30,253){\line(0,-1){50}}
\multiput(30,198)(0,-10){2}{\line(0,-1){5}}\put(30,178){\vector(0,-1){15}}
\put(250,5){ ${W'''}$}
\put(260,245){${W'}$}
\put(0,0){${ W}$}
\put(0,255){${W''}$}
\put(45,115){$a_4$}\put(-23,150){$c_3$}\put(65,20){$b_3$}\put(135,-20){$c_2$}
\put(220,20){$a_1$}\put(150,60){$c_5$}\put(130,153){$z$}\put(290,133){$c_6$}
\put(202,250){$a_3$}\put(130,295){$c_4$}\put(55,255){$b_1$}\put(65,195){$a_2$}
\put(220,155){$b_4$}\put(205,83){$b_2$}\put(115,215){$c_1$}
\end{picture}}
\put(400,100){\begin{picture}(300,300)
\thicklines
\multiput(141,0)(60,100){2}{\line(2,1){80}}
\multiput(140,0)(60,100){2}{\line(2,1){80}}
\multiput(139,0)(60,100){2}{\line(2,1){80}}
\multiput(140,1)(80,40){2}{\line(3,5){60}}
\multiput(140,0)(80,40){2}{\line(3,5){60}}
\multiput(140,-1)(80,40){2}{\line(3,5){60}}
\multiput(0,141)(60,-40){2}{\line(3,5){60}}
\multiput(0,140)(60,-40){2}{\line(3,5){60}}
\multiput(0,139)(60,-40){2}{\line(3,5){60}}
\multiput(1,140)(60,100){2}{\line(3,-2){60}}
\multiput(0,140)(60,100){2}{\line(3,-2){60}}
\multiput(-1,140)(60,100){2}{\line(3,-2){60}}
\multiput(0,141)(60,-40){2}{\line(4,-5){80}}
\multiput(0,140)(60,-40){2}{\line(4,-5){80}}
\multiput(0,139)(60,-40){2}{\line(4,-5){80}}
\multiput(200,101)(80,40){2}{\line(-4,5){80}}
\multiput(200,100)(80,40){2}{\line(-4,5){80}}
\multiput(200, 99)(80,40){2}{\line(-4,5){80}}
\multiput(61,240)(60,-40){2}{\line(2,1){80}}
\multiput(60,240)(60,-40){2}{\line(2,1){80}}
\multiput(59,240)(60,-40){2}{\line(2,1){80}}
\multiput(141,0)(60,240){2}{\line(-3,2){60}}
\multiput(140,0)(60,240){2}{\line(-3,2){60}}
\multiput(139,0)(60,240){2}{\line(-3,2){60}}
\multiput(160,80)(30,-20){2}{\line(3,-2){20}}
\multiput(220,180)(30,-20){2}{\line(3,-2){20}}
\multiput(80,40)(30,15){3}{\line(2,1){20}}
\multiput(0,140)(30,15){3}{\line(2,1){20}}
\multiput(160,80)(-20,25){4}{\line(-4,5){16}}
\multiput(220,180)(-20,25){4}{\line(-4,5){16}}
\multiput(220,180)(-16,-26.667){4}{\line(-3,-5){12}}
\multiput(140,280)(-16,-26.667){4}{\line(-3,-5){12}}
\multiput(140,140)(-30,20){2}{\line(-3,2){20}}
\multiput(140,140)(-30,-15){3}{\line(-2,-1){20}}
\multiput(140,140)(20,-25){4}{\line(4,-5){16}}
\multiput(140,140)(16,26.667){4}{\line(3,5){12}}
\multiput(140,0)(60,100){2}{\circle*{8}}
\multiput(220,40)(60,100){2}{\circle*{8}}
\multiput(60,100)(60,100){2}{\circle*{8}}
\multiput(0,140)(60,100){2}{\circle*{8}}
\multiput(140,280)(80,-100){2}{\circle*{8}}
\multiput(80,180)(80,-100){2}{\circle*{8}}
\multiput(80,40)(120,200){2}{\circle*{8}}\put(140,140){\circle*{12}}
\put(270,0){\sc${W''}$}
\put(-10,-20){\sc${W'}$}
\put(260,265){\sc${W}$}
\put(20,275){\sc${W'''}$}
\put(260,20){\vector(-1,1){50}}\put(250,260){\line(-1,-1){42}}
\put(203,213){\vector(-1,-1){20}}
\put(16,6){\vector(1,1){63.5}}\put(38,260){\vector(0,-1){80}}
\put(45,115){$a_4$}\put(-23,150){$c_3$}\put(65,20){$b_3$}\put(135,-20){$c_2$}
\put(220,20){$a_1$}\put(150,60){$c_5$}\put(130,153){$z$}\put(290,133){$c_6$}
\put(202,250){$a_3$}\put(130,295){$c_4$}\put(55,255){$b_1$}\put(65,195){$a_2$}
\put(220,155){$b_4$}\put(205,83){$b_2$}\put(115,215){$c_1$}
\end{picture}}
\put(327,210){\huge\bf $=$} 
\end{picture}
\caption{ Graphical representation for the tetrahedron
  equations for interaction-round-a-cube models.}\label{TEIRC}
\end{figure}
It satisfies the tetrahedron
equation in the ``interaction-round-a-cube'' form 
\cite{Baxter:1986phd} (see Fig.~\ref{TEIRC}),
\be
\begin{array}{l}
\ds \int_{\mathbb{R}} dz\, W(a_4|c_1,c_3,c_2|b_3,b_2,b_1|z)\,
W'(c_1|a_3,b_1,b_2|z,c_6,c_4|b_4)\\[.3cm]
\qquad\times W''(b_1|c_4,c_3,z|b_3,b_4,a_2|c_5)\,
W'''(z|b_4,b_3,b_2|c_2,c_6,c_5|a_1)=\\[.4cm]
\phantom{XXXX} \ds= \int_{\mathbb{R}} dz
\,W'''(b_1|c_4,c_3,c_1,|a_4,a_3,a_2|z)\,
W''(c_1|a_3,a_4,b_2|c_2,c_6,z|a_1)\\[.3cm]
\phantom{XXXXXX} \qquad \times W'(a_4|z,c_3,c_2|b_3,a_1,a_2|c_5)\,
W(z|a_3,a_2,a_1|c_5,c_6,c_4|b_4),\label{tetr}
\end{array}
\ee where the four sets of the spectral parameters are constrained
as
\begin{equation}
\T_1'=\T_1,\quad \T_1''=-\T_2,\quad \T_1'''=\T_3, \quad
\T_2''=\T_2',\quad \T_2'''=-\T_3', \quad
\T_3'''=\T_3''\ .\label{Tshki}
\end{equation}
Note that the parameters $\T_1,\T_2,\T_3$ are similar to those in the 
Zamolodchikov model \cite{Zamolodchikov:1981kf} and its
generalization for an arbitrary number of spin states
\cite{Bazhanov:1992jq}, which is 
considered in Sect.~\ref{cyc-sol} below.
One can relate them as $\T_j=\log[\tan(\theta_j/2)]$ 
to the angles $\theta_1,\theta_2,\theta_3$ of the spherical triangle in
Sect.~\ref{cycrep-def}. In total,
there are twelve parameters
$\T_j,\T_j',\T_j'',\T_j'''$, $j=1,2,3$
(three for each weight function)  constrained by six relations 
\eqref{Tshki}. Thus, the tetrahedron equation
\eqref{tetr} contains six independent parameters (in contract 
to only five parameters in the cyclic case; see the text after
Eq.\eqref{six-angles} below). 
 
The above solution can be simply generalized 
by multiplying the weight \eqref{Wweight} by an ``external field''
factor 
\be
W(\ldots)\to \exp\Big[{\sum_{j=1}^3 f_j (\sigma_j+\sigma_j')}\Big]
\,W(\ldots)\ .
\ee
This substitution does not affect the validity of the tetrahedron
equation  \eqref{tetr}, provided the field parameters $f_j$ for
different $W$'s are constrained as
\begin{equation}
f_3''=f_3'-f_3,\quad f_1'''=f_1''-f_1', \quad
f_2'''=f_2''+f_1,\quad f_3'''=f_2-f_2'.\label{ashki}
\end{equation}
Similar generalizations apply for the solutions \eqref{R-fock} and
\eqref{R-modular}. The solution \eqref{Wweight} was previously
obtained in \cite{BMS08a}.

\subsection{Cyclic representation solution}\label{cyc-sol}
\subsubsection{Generalized form of the $q$-oscillator map} 
Here, we will solve equation \eqref{rl-def} with a more general form
of the map \eqref{themap}, considered in \cite{BS05}. 
Let
\be\label{loper}
{\mathcal L}({\mathcal H}_j)=
\left(\begin{array}{cccc} 1 & 0 & 0 & 0 \\
0 & \lambda_j \kop_j & \bos_j^+ & 0 \\ 0 & \lambda_j\mu_j\bos_j^-
& -\mu_j\kop_j & 0 \\ 0 & 0 & 0 &
\lambda_j\mu_j\end{array}\right)\qquad j=1,2,3,
\ee
be an operator-valued matrix acting in the direct product of two
vector spaces ${\mathbb C}^2\otimes{\mathbb
  C}^2$, whose elements depends on a set of generators
${\mathcal H}_j=\{k_j,a_j,a^*_j\}$ of the $q$-oscillator algebra
\eqref{q-osc1} and
continuous parameters $\lambda_j$, and $\mu_j$, where $j=1,2,3$. 
The RHS of \eqref{loper} is understood as a block matrix with
two-dimensional blocks where matrix indices of the second space ${\mathbb
  C}^2$ numerate the blocks, while those for first space numerate
the elements inside the blocks. 

The required generalization of \eqref{rl-def} reads
\cite{BS05}\footnote{Equation \eqref{TE-aux} reduces to \eqref{rl-def}
  when $\lambda_1=\lambda_2=\lambda_3=1$ and $\mu_1=\mu_2=\mu_3=-\varepsilon$,
  where $\varepsilon$ enters the map \eqref{themap}.}
\begin{equation}\label{TE-aux}
L_{12}({\mathcal H}_1)\, L_{13}({\mathcal H}_2)
L_{23}\,({\mathcal H}_3)=R_{123}\,
L_{23}({\mathcal H}_3)\,L_{13}({\mathcal H}_2)\,
L_{12}({\mathcal H}_1) \, R_{123}^{-1}
\end{equation}
Here ${\mathcal L}_{12}$  denote the matrix which 
acts as \eqref{loper} in the first and second component of the
tensor product ${\mathbb C}^2\otimes{\mathbb
  C}^2\otimes{\mathbb C}^2$ and coincides with the unit operator in
the third component (${\mathcal L}_{13}$ and ${\mathcal L}_{23}$ are
defined similarly). It is important to note that the 
additional parameters $\lambda_j$ and $\mu_j$ enter \eqref{TE-aux} only
in three combinations $\lambda_2/\lambda_3$, $\lambda_1\mu_3$ and 
$\mu_1/\mu_2$ (explicit formulae for the corresponding 
generalization of the map
\eqref{themap} are given in Eqs.(24-26) of \cite{BS05}).

\subsubsection{Cyclic representations of the $q$-oscillator algebra}
\label{cycrep-def}  
Let $N\ge3$ be an odd integer,
\begin{equation}\label{cyclic-q}
q=-\EXP^{i\pi/N},\quad q^N=1,\quad N=\mbox{odd},\quad N\ge3.
\end{equation}
and $V^N$ be an $N$-dimensional vector
space, spanned by the vectors $|n\rangle$, \ $n\in{\mathbb
  Z}_N=\{0,1,\ldots,N-1\}$. Define $N$ by $N$ matrices 
(the index $n$ is treated {\em modulo} $N$) 
\begin{equation}
X\,|n\rangle = q^n\,|n\rangle,\quad \quad Z\,|n\rangle =
|n+1\rangle,\qquad X^N=Z^N=1,
\end{equation}
and their embedding into a direct product $V^N\otimes V^N\otimes V^N$,
\be
X_j=1\otimes\cdots\otimes \mathop{X}_{j\mbox{-th}}\otimes\cdots\otimes1,\quad
Z_j=1\otimes\cdots\otimes \mathop{Z}_{j\mbox{-th}}\otimes\cdots\otimes1,
\quad j=1,2,3. 
\ee
Equation \eqref{TE-aux} involves the direct product of  
three copies of the $q$-oscillator algebra \eqref{q-osc1},
labeled by the subscript $j=1,2,3$. The most general cyclic representation 
for this product 
\begin{equation}\label{cyc-rep}
\kop_j=\varkappa_j\,X_j\;,\quad
\bos_j^*=\frac{1}{\rho_j}(1-q^{-1}\varkappa_j^2\,X^2_j)\,Z_j\;,\quad
\bos_j\phantom{{}^*}=\rho_j\,Z_j^{-1}, \quad j=1,2,3, 
\end{equation}
contains two continuous parameters $\varkappa_j,\rho_j$, for 
each factor, so there six arbitrary parameters altogether.
Detailed inspection of \eqref{TE-aux} shows, however, that  
$\rho_1,\rho_2,\rho_3$ only enter through a ratio
$\rho_1\rho_3/\rho_2$, so there are only four essential parameters. 

Let
$\theta_1, \theta_2, \theta_3$ and $a_1,a_2,a_3$ be angles and sides 
of a spherical triangle, respectively. Define
\be
2\beta_1=a_2+a_3-a_1,\quad 2\beta_2=a_1+a_3-a_2,\quad
2\beta_3=a_1+a_2-a_3,\quad \beta_0=\pi-\beta_1-\beta_2-\beta_3.
\ee
and set
\be
\varkappa_1=\ii\sqrt[N]{\tan\frac{\theta_1}{2}},\quad
\varkappa_2=\ii\sqrt[N]{\cot\frac{\theta_2}{2}},\quad
\varkappa_3=\ii\sqrt[N]{\tan\frac{\theta_3}{2}}\ .
\end{equation}
Then Eq.\eqref{TE-aux} implies 
\begin{equation}
{\rho_1\rho_3}/{\rho_2}\;=\;\EXP^{-\ii\beta_2/N}\sqrt[N]{{\sin
a_2}/{\sin\beta_0}},
\ee
and 
\begin{equation}
{\lambda_3}/{\lambda_2}=\EXP^{\ii a_1/N}\;,\quad
\lambda_1\mu_3=\EXP^{-\ii a_2/N}\;,\quad
{\mu_1}/{\mu_2}=\EXP^{\ii a_3/N}\;.
\end{equation}

\subsubsection{Solution of the tetrahedron equation}
Let $p$ denote a point on Fermat curve 
\be \label{Fermat-def}
p\opr
(x,y\;|\;x^N+y^N=1). 
\ee 
Introduce a function $\cdil_p(n)$,
$n\in{\mathbb Z}_N$, (cyclic analog of the quantum dilogarithm)
\cite{Bazhanov:1992jq}
\begin{equation}\label{W-Fermat}
\cdil_p(0)=1\;,\quad
\frac{\cdil_p(n-1)}{\cdil_p(n)}\;=\;\frac{1-xq^{2n}}{y},\quad
\cdil_p(n+N)=\cdil_p(n)
\end{equation}
The $R$-matrix solving \eqref{TE-aux} for the cyclic representation
\eqref{cyc-rep} 
is given by
\begin{equation}\label{R-cyclic}
\begin{array}{l}
\ds \langle n|R|n'\rangle \;=\;
\delta_{n_1+n_2,n_1'+n_2'}
\delta_{n_2+n_3,n_2'+n_3'}\times\phantom{XXXXXXXXXXXXXXX}\\
[2mm]
\phantom{xxxxxxxx} \ds
q^{n_1n_3-n_2'(n_1+n_3)}\sum_{n\in\mathbb{Z}_N}\;q^{-2nn_2'}\;
\frac{\cdil_{p_1}(n+n_1+n_3')\cdil_{p_2}(n)}{\cdil_{p_3}(n+n_1)\cdil_{p_4}(n+n_3)}\;,
\end{array}
\end{equation}
where the points $p_j$ on the curve \eqref{Fermat-def} are defined by
\begin{equation}\label{O-point-spherical}
\begin{array}{ll}
\ds  x_1=\EXP^{-\ii a_2/N}
\sqrt[N]{{\sin\beta_2}/{\sin\beta_0}}\;, &\quad \ds
y_1=\EXP^{\ii\beta_2/N}\sqrt[N]{{\sin
    a_2}/{\sin\beta_0}}\;,\\[.3cm]
\ds x_2=\EXP^{-\ii a_2/N}
\sqrt[N]{{\sin\beta_0}/{\sin\beta_2}}\;, &\quad
\ds y_2=\EXP^{\ii\beta_0/N}\sqrt[N]{{\sin a_2}/{\sin\beta_2}}\;,\\[.3cm]
\ds x_3= \EXP^{- \ii (a_2+\pi)/N}
\sqrt[N]{{\sin\beta_3}/{\sin\beta_1}}\;, &\quad\ds
y_3=\EXP^{-\ii\beta_3/N}\sqrt[N]{{\sin
a_2}/{\sin\beta_1}}\;,\\[.3cm]
\ds x_4= \EXP^{-\ii (a_2+\pi)/N}
\sqrt[N]{{\sin\beta_1}/{\sin\beta_3}}\;, & \quad\ds
y_4=\EXP^{-\ii\beta_1/N}\sqrt[N]{{\sin a_2}/{\sin\beta_3}}\;.
\end{array}
\end{equation}
We write this $R$-matrix as
$R_{123}(\theta_1,\theta_2,\theta_3)$ to indicate its dependence on
the angles $\theta_1,\theta_2,\theta_3$. 
It satisfies the tetrahedron equation \eqref{TE},
provided the $R$-matrices therein are parametrized by four triples of 
dihedral angles at four vertices of an Euclidean tetrahedron 
\begin{equation}\label{six-angles}
\begin{array}{ll}
R_{123}=R_{123}(\theta_1,\theta_2,\theta_3),&\qquad
R_{145}=R_{145}(\theta_1,\pi-\theta_4,\pi-\theta_5),\\[.3cm]
R_{246}=R_{246}(\pi-\theta_2,\pi-\theta_4,\theta_6),&\qquad
R_{356}=R_{356}(\theta_3,\theta_5,\theta_4)\ .
\end{array}
\end{equation}
Here $\theta_j$, $j=1,2,\ldots,6$ are \emph{inner} dihedral angles of 
the tetrahedron.  These angles are constrained by one relation, therefore
Eq.\eqref{TE} in this case contains five arbitrary parameters.  
Finally, the substitution 
\begin{equation}
\begin{array}{lll}
n_1=g+f-a-b,&\quad n_2=a+c-e-g,&\quad
n_3=e+f-a-d,\\[.2cm]
 n_1'=c+d-e-h,&\quad
n_2'=f+h-b-d,&\quad n_3'=b+c-g-h
\end{array}
\end{equation}
and some trivial equivalence transformations bring (\ref{R-cyclic}) 
into its ``interaction-round-a-cube'' form     
\begin{equation}\label{W-cyclic}
W(a|e,f,g|b,c,d|h)\;=\;\sum_{n\in\mathbb{Z}_N} q^{2n(b+d-f-h)}
\frac{\cdil_{p_1}(n-h+c)\cdil_{p_2}(n-f+a)}{\cdil_{p_3}(n-b+g)
\cdil_{p_4}(n-d+e)}\;,
\end{equation}
which satisfies (\ref{tetr}) where the integration
is replaced by the summation over $\mathbb{Z}_N$, while the angle
parametrization remains the same as in \eqref{six-angles}. 
Although our considerations in the cyclic case were restricted to odd
values of $N$ in \eqref{cyclic-q}, the final result \eqref{W-cyclic}
as valid for even values of $N$ as well. The solution \eqref{W-cyclic}
was previously obtained in \cite{Bazhanov:1992jq}; for $N=2$ it
coincides with the Baxter's form \cite{Baxter:1986phd} for the
Boltzmann weights the Zamolodchikov model \cite{Zamolodchikov:1981kf}. 

\section{Conclusion}
In this paper we have exposed various connections between
discrete differential geometry and  statistical mechanics,
displaying geometric origins of algebraic structures
underlying integrability of quantum systems.

We have shown that the 3D circular lattices are associated with an
integrable discrete Hamiltonian system and constructed three different
quantizations of this system.  In this way in Sect.~\ref{solutions}
we have obtained all
previously known solutions of the tetrahedron equation. They are given
by Eqs. \eqref{R-fock}, \eqref{R-modular}, \eqref{Wweight}, \eqref{R-cyclic}.
The resulting 3D integrable models
can be thought of as describing quantum fluctuations of the lattice
geometry. The classical geometry of the 3D circular lattices reveals
itself \cite{BMS08a} as
a stationary configuration giving the leading contribution to the
partition function of the quantum system in the quasi-classical
limit.

The solutions of the tetrahedron equation discussed here possess
a remarkable property: they can be used to obtain 
\cite{Bazhanov:1992jq,BS05} infinite number of  
two-dimensional solvable models related to
various representations of
quantized affine algebras $U_q(\widehat{sl}_n)$,
$n=2,3,\ldots,\infty$ (where $n$ coincides with the size
of the ``hidden third dimension'').
Apparently, a similar 3D interpretation, originating from other
simple geometrical models, also exists for the
trigonometric solutions of the Yang-Baxter equation, related with all
other infinite series of quantized affine algebras
\cite{Bazhanov:1984gu,Jimbo86} and super-algebras
\cite{Bazhanov:1987tmp} (see \cite{Ser09a,Ser09b} for recent results
in this direction). Therefore, it might very well
be that not only the phenomenon of quantum
integrability but the quantized algebras
themselves are deeply connected with geometry.

There are many important outstanding questions remain, in
particular, the geometric meaning of the Poisson algebra
\eqref{poisson} and connections of the 3D circular lattices with the 2D
circle patterns \cite{BSp} on the plane or the sphere.
It would also be interesting to understand underlying reasons of a
``persistent'' appearance of the $q$-oscillator algebra \eqref{q-osc1}
as a primary algebraic structure in many other important aspects
of the theory of integrable systems, such as, for example, the
construction of Baxter's ${\bf Q}$-operators \cite{BLZ97a} and
the calculation of correlation functions of the XXZ model \cite{BJMS06}.
It would also be
interesting to explore connections of our results
with the invariants of the 3D manifolds \cite{W89,RT91,TV92},
the link invariants \cite{Kash95,Mur2001,Hikami:2006},
quantization of the Techmueller space \cite{Kash98,Fok99} and
the representation theory of $U_q(sl(2|{\mathbb R})$ \cite{PT01}.
So, there are many interesting questions about the quantum integrability
still remain unanswered, but one thing is getting
is more and more clear: it is not just connected with geometry, it is
geometry itself! (though the Quantum Geometry).

\section*{Acknowledgments}
The authors thank R.J.Baxter, M.T.Batchelor, A.Doliwa, M.Jimbo, R.M.Kashaev,
T.Miwa, F.A.Smirnov, Yu.B. Suris and P. Vassiliou 
for interesting discussions and remarks.
One of us (VB) thanks M.Staudacher for his hospitality at the
Albert Einstein
Institute for Gravitational Physics in Golm, where some parts of
this work have been done. Special thanks to A.I.Bobenko for numerous
important comments and to D. Whitehouse at the ANU Supercomputer Facility
for the professional graphics of the Miquel circles (Fig.~7).



\newcommand\oneletter[1]{#1}

\end{document}